**Proposed experiments to clarify the real nature of the quantum waves**

J. R. Croca[1,2], P. Castro[2 a], M. Gatta[2,3] and R. N. Moreira[2]

University of Lisbon, Faculty of Sciences,

Department of Physics[1] and the Center for the Philosophy of Sciences of the University of Lisbon (CFCUL) [2]

CINAV and Escola Naval (Portuguese Naval Academy)[3]

Abstract: The nature of quantum waves, whether they are real physical waves or, on the contrary, mere probability waves, has been a very controversial theme since the beginning of quantum theory. Here we present some possible experiments that may clarify the problem.

**1 - Introduction**

The development of hydrodynamic quantum analogs (HQA) experiments, using droplets [1], have shown that pilot wave phenomena do exist in nature, at least at the macroscopic level. Furthermore, experimenters from fluid dynamics have used HQA to study several cases, where an analogy between hydrodynamic and quantum phenomena can be established, according to de Broglie Pilot wave theory hypothesis in quantum mechanics. This realization and the recent proposal of an experimental setup designed to detect quantum waves [2], has attracted the interest of a small community of physicists and philosophers of science [3, 4] about the ontic nature of these waves. The present work is an extension of a previous one [2] proposing a set

---

[a] paulo.castro.pi@gmail.com

of improved experiments, designed to obtain evidence of the existence of quantum waves.

In the Copenhagen conceptual framework, quantum waves are mere mathematical probability waves, therefore devoid of any physical reality. On the other hand, de Broglie and his school tradition maintain that quantum waves do have some physical reality. That is, they stand for something existing independently of the observer. Furthermore, the physical entity usually named in the Copenhagen tradition the "quantum particle" is accepted in de Broglie's approach to be a complex nonlinear composition of a pure real wave, which can be called $\theta$, practically devoid of energy, and a relatively high energetic kernel, which can be called $\xi$, identifiable with a "corpuscle" or a "singularity" and that corresponds to what is usually detected in standard quantum experiments. The corpuscle $\xi$, is thus surrounded by its pilot wave $\theta$, as represented schematically below in Fig.1.1

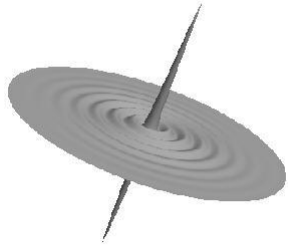

Fig.1.1 - Graphic representation of a quantum particle, according to de Broglie school tradition

The usual detection processes occur due to the strong interaction of the corpuscle $\xi$, with common detectors, since its pilot wave $\theta$, has in principle practically no energy, when compared to the corpuscle. Hence, resulting that the currently used detectors are unable to detect it. To obviate this problem a detection strategy can be adopted by the use of the so called "empty quantum waves". These are quantum waves without any corpuscle, that in principle can be harnessed, simply by considering the beamsplitter output branch along which the corpuscle has not followed.

The pilot wave, as the name indicates, "guides" or "pilots" the corpuscle into the regions where the extended $\theta$ pilot wave is more intense. Consequently, the probability of finding the corpuscle is proportional to the density of the pilot wave, as required by Born's rule:

$$p(r) \propto |\theta|^2, \qquad (1.1)$$

Something that was proposed, very early on, by de Broglie [5].

The quest for de Broglie's concept of empty waves, although unpopular, has nevertheless been discussed in a number of previous works, as in [6] and also in [7, especially section 8.8] and [8]. The conceptual principles beside these experiments, concerning the nature of quantum waves, were developed mainly since the seventies by researchers directly associated with Louis de Broglie [6], namely, J. and M. Andrade e Silva [9], F. Selleri [10], A. Garuccio, V. Rapisarda and J.P. Vigier [11], J. R. Croca, R. Moreira [12, 13] and P. Neves [14] (these ones associated to Andrade e Silva). With this work we hope to contribute to a further clarification of the problem of discerning the ontological nature of quantum waves, if whether they are real physical waves or mere probability waves.

**2 - Proposed experiments to detect empty quantum waves**

If, on one hand, we believe that the problem of the existence of the so-called empty waves or de Broglie waves, has not been experimentally settled, beyond any reasonable doubt [2], on the other, we also concede that there are technical objections, concerning the feasibility of such detection. In this work we have focused on two possible arguments that can be invoked to dismiss the claim that quantum waves exist. The first is the bosonic effect, or the bunching effect, related with the so-called "Mandel dip", which although concerning the photons behavior, may also be hypothesized to apply to photonic quantum waves. Thus, dismissing the conclusion in photon emission experiments that a quantum wave exists, since it has not reached the proper detecting apparatus, due to that bosonic effect.

The Mandel dip, or perhaps better, the Hong-Ou-Mandel effect [15] is due to a kind of "bosonic effect" happening between two photons. The phenomena can be described in the following way. For a temporal overlap between the two photons ("signal" and "idler") coming out from a beamsplitter, both having been emitted from the same parametric-down-converter, it is observed that both photons bunch together, so that either both reach a detector D1, or both reach detector D2, along the orthogonal arms, for equal optical path-lengths. Consequently, due to the symmetry of the apparatus, both detectors D1 and D2 read the same number of counts, that is to say, 50%, of the total counting. Nevertheless, an imposed asymmetry in both optical pathlengths (simply increasing the length of one of the arms) will destroy the otherwise equal distributions of detections. The more so, the larger the deviation of one arm pathlength from the other, thus making possible the observation of the so-called "Mandel dips". For the symmetric situation (that is, for equal arms pathlengths), a counting-rate base line will form as represented by the thick dotted line in Fig.3.4, ahead.

One of our main objectives in the optical quantum experiments presented herein, was to prevent the attribution of unequal distributions of counts (and therefore of Mandel dips) to the asymmetry in the optical pathlengths. This aim was carefully taken into consideration in the design of the experimental setups for quantum wave detection.

Let us now discuss the second possible argument, dismissing the detection of quantum waves. In recent years, optical quantum experiments have been performed [16, 17, 18] that, in our view, suggest the physical reality of quantum waves [19], even though their authors argue that the results are still explainable within the Copenhagen school framework.

In fact, it is precisely from the argumentation used in these works [16, 17, 18] that we have considered the second possible objection against the empirical validity of empty quantum wave detections, in the proposed experiments. This objection concerns the claim that two twin photons, produced in the same parametric down converter crystal, are entangled, and as such, behave as a "single indivisible entity", that is, as a kind of "bi-photon", as already suggested by Shih [20]. Using this bi-photon picture, one can thus say that the time and phase correlation between two photons, that were produced in the same parametric down converter crystal, upon a detection at the end of their journeys, is due to the fact that they are a single entity. In other words, there are really no two photons in mid-flight, but rather a single quantum entity, regardless of the physical distance between the arms of the interferometer. And in such a way that their observed quantum states, upon the final detection, including their polarizations and phases, would be correlated "instantaneously", without need for another entity. This would be, of course, contrary to our own proposal, where a de Broglie empty wave, would have to be physically present to justify the observed correlation between the photons. Preventing this bi-photon effect interpretation of the results was the other of our two aims, designing the experiments below.

We now introduce a set of experiments designed to detect the existence of empty $\theta$ waves, that is, pilot waves that have been separated from their guided corpuscles. The experiments were developed, with several versions progressively coping with the possible wave bosonic effect and the bi-photon effect. In what follows we will use the expressions "de Broglie wave", "pilot wave" and "guiding wave" with the same meaning. We also use the greek letter $\phi$ to denote a de Broglie wave where a corpuscle is present and the greek letter $\theta$ to denote a de Broglie empty wave. A line connecting different detectors in the diagrams below means that the detectors are in coincidence.

## 2.1 - Three slits experiments

The first experiment, shown in Fig.2.1, is basically composed of a nonlinear crystal plus a screen with three slits and two detectors, $D_R$ and D, put (electronically) in coincidence, so that detector D is "on" only when $D_R$ is excited by an incoming corpuscle. The dotted lines represent two equal beamsplitters, in a symmetric configuration, allowing for equal average intensities from the two beams.

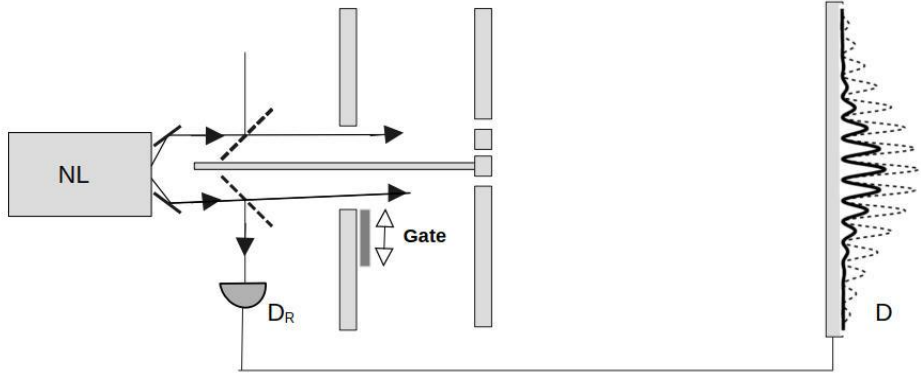

Fig.2.1 - Three slits experiment

The transmitted signal photon emitted by the nonlinear crystal in a parametric down conversion goes, independently of the idler photon, to the screen, only interacting with two of the slits along its way (the upper branch in Fig.2.1, is physically separated from the lower slit).

For the behavior of the idler photon, in particular, there are two possibilities:

a) According to the usual ("Copenhagen") approach, the idler photon may be reflected or transmitted at the lower beamsplitter.

If reflected, it activates the detector $D_R$ that is connected in a coincidence setup to the large detector, D, placed in the far field plane relative to the slits. In such a setup, any idler photon can be detected by the large detector D, not only after reflection, but also directly, by transmission, at the lower beamsplitter. After detection, nothing remains in the device due to the collapse of the wave function.

b) Following the de Broglie's approach, the idler photon guiding wave is splitted with both reflected, $\theta_r$, and transmitted, $\theta_t$, components, while the corpuscle is either reflected or else transmitted. In the case that the corpuscle is reflected, it interacts with the detector $D_R$. Still, along the transmitted path it should follow, we submit, a transmitted empty pilot wave, $\theta_t$, towards the lower slit. This wave could be blocked at will by the action of a gate.

Let us now see the predictions for the overall behavior of the device:

a) Usual ("Copenhagen") method:

Since, by construction, the large array of detectors at D is activated only when the reduction detector $D_R$ is excited, this means that not only nothing comes out from the third slit, but also that the expected distribution of counts at D will be given, as usual, by

$$I^U \propto |\psi_1 + \psi_2|^2 \qquad (2.1)$$

going through the two upper slits, giving

$$I^U \propto (1 + \cos\varphi), \qquad (2.2)$$

in which $I^U$ is the intensity in the usual interpretation and $\varphi$ represents the constant phase difference between the two waves, assuming that

$$|\psi_1| = |\psi_2|. \tag{2.3}$$

The fringe visibility given by the traditional Born-Wolf rule, $V = (I_M - I_m)/(I_M + I_m)$, is 1, $V^U = 1$, with $I_M = 1 + 1$ and $I_m = 0$.

That is, after a certain time, a clear interferometric pattern will appear at the large detector D.

b) de Broglie's approach:

According to de Broglie way of thinking, to predict the expected counting distribution $I^B$ at the detector D, we have to consider not only the two waves $\phi_1$ and $\phi_2$ from slit one and two, but also the empty theta wave, $\theta$, coming from the third slit, so that we have:

$$I^B \propto |\phi_1 + \phi_2 + \theta|^2. \tag{2.4}$$

Again, assuming equal wave amplitudes for all three waves, and that the first two waves are coherent with each other and incoherent in relation to the third wave $\theta$, we have, after some simple calculations [6]:

$$I^B \propto \left(1 + \frac{2}{3}\cos\varphi\right), \tag{2.5}$$

giving for the fringe visibility the value $V^B = 2/3$, with $I_M = 1 + 2/3$ and $I_m = 1 - 2/3$.

Summarizing:

For this experiment, the usual ("Copenhagen") approach predicts a clear interference pattern with visibility one.

The de Broglie's approach predicts a blurring in the interferometric pattern. That is to say, a change of visibility from 1 to 2/3.

The experimental protocol must hence be followed along two steps. Observations must be made:

1 - With the gate closed, so that no empty pilot wave can reach the third slit.

2 - With the gate opened, allowing the empty pilot wave to reach the third slit and interfere with the other two waves, producing the blur in an otherwise clear interferometric pattern.

For this experiment, it must nevertheless be said that, at the lower branch of the NL source, it may happen that a kind of bosonic effect applies to the quantum waves, rendering null the empty quantum wave detection. This happens because if $D_R$ has been triggered, then it also means that both quantum waves, the one with the triggering corpuscle and the other, the empty quantum wave, have not been splitted by the beamsplitter. As such, both waves end up at $D_R$ and no empty wave has progressed through the third slit below. The other problem that may arise is the one concerning the possible bi-photon effect. Since both photons coming from the same NL source may be entangled, this means that they may act as single quantum entity. Consequently, according to "Copenhagen", the detection at $D_R$ produces an overall collapse of all quantum wave branching and it is this that explains the absence of an interference pattern at the screen D, again denying the existence of an empty quantum wave.

In the next experimental version, we deal first with the bi-photon effect problem.

### 2.1.1 - Three slits experiment variant to overcome the eventual bi-photon effect

This variant of the previous experiment, shown in Fig.2.2, aims to avoid an eventual possible entangled "bi-photon" effect between the two generated photons. That is to say, a possible phase correlation between the lower photon and the upper photon, a condition that, for an empty wave quantum detection, would turn discardable the prediction (2.5), above.

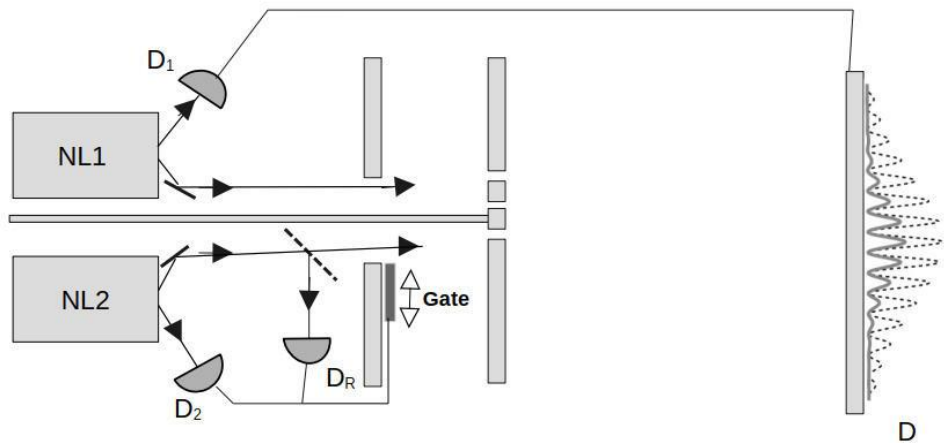

Fig.2.2 - Three slits experiment variant to overcome a possible bi-photon effect

In this experimental setup, the completely independent sources are two nonlinear crystals. Detectors $D_1$ and $D_2$ are put in coincidence and used to ensure that NL1 and NL2 are monophotonic sources.

As in the previous experiment, when the gate is closed, we expect an interference pattern with visibility 1 to appear.

If the gate is opened by the action of the $D_R$ detector in this setup, an empty wave may then follow through the third slit, overlapping the other two waves. Since the two monophotonic sources are completely independent, their emission of photons is random. However, it may happen that, sometimes, the two independent photons may be emitted at about the same time. In such condition, we may have a total or partial superposition at the screen and consequently some blurring, now and then, of the interference pattern predicted by the usual approach. These expected differences in the fringe visibility may be increased if one raises the relative emission rate of the second source, NL2.

## 3 - Mach-Zehnder interferometer experiments to detect empty quantum waves

In essence, the three slit experiments above and the experiments done with the Mach-Zehnder interferometer below are similar. The difference laying in the precision that favors this particular interferometric technique.

Consider the experimental sketch in Fig.3.1.

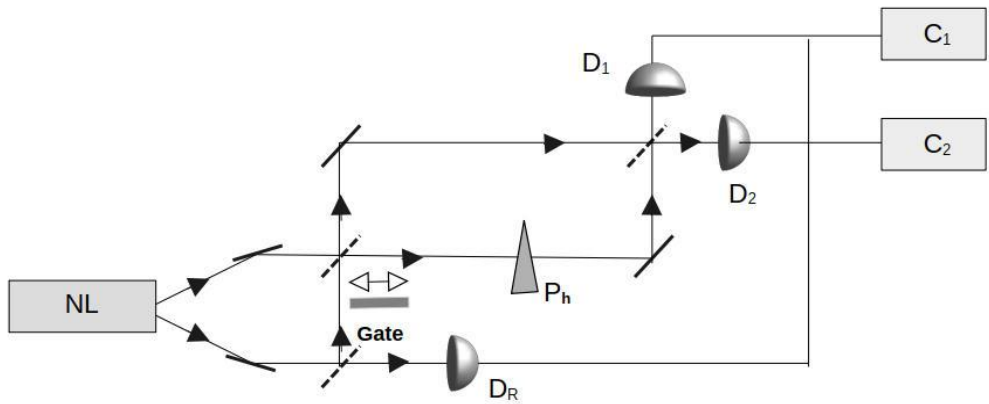

Fig.3.1 - Mach-Zehnder Interferometer. NL = nonlinear crystal, DD = detectors, CC = counters, Ph = phase shifter

Let us write the predictions for this experimental setup, according to the two approaches that we have been considering.

a) Usual ("Copenhagen") approach:

After the reduction detector $D_R$ clicks, nothing enters the interferometer. So, the expected results are the same whether the gate is open or closed and the intensities registered at the two outputs C1 and C2 are [8]:

$$\begin{cases} I_1 = \frac{1}{2}I_0(1 - \cos \delta_\phi) \\ I_2 = \frac{1}{2}I_0(1 + \cos \delta_\phi) \end{cases} \quad (3.1)$$

where $\delta_\phi$ represents the relative phase difference of the waves entering the interferometer from the signal source. By calibrating the interferometer with the help of a phase shifting device, $P_h$, so that $\delta_\phi = 0$, we have:

$$\begin{cases} I_1 = 0 \\ I_2 = I_0 \end{cases} \quad (3.2)$$

b) de Broglie's approach:

b1) Gate closed.

In this situation the predictions are the same as the usual ("Copenhagen") approach.

b2) Gate open.

In this case the results are expected to be quite different from the "Copenhagen" approach, due to the interference effect of the $\theta$ waves at the final part of the Mach-Zehnder interferometer. In fact, the predicted results must also include the empty waves, resulting:

$$\begin{cases} I_1 \propto |\phi_{TR} + \phi_{RT} + \theta_{TT} + \theta_{RR}|^2 \\ I_2 \propto |\phi_{RR} + \phi_{TT} + \theta_{RT} + \theta_{TR}|^2 \end{cases} \quad (3.3)$$

Assuming, for simplicity, that all mixing quantum waves have equal intensity, we have

$$|\theta_{RT}|^2 = |\theta_{TR}|^2 = |\theta_{TT}|^2 = |\theta_{RR}|^2 = \frac{1}{4}|\theta|^2 \quad (3.4)$$

giving, after some easy calculations [8]:

$$\begin{cases} I_1 = \frac{1}{2}I_0(1 + \cos\delta_\phi - \cos\delta_\theta) \\ I_2 = \frac{1}{2}I_0(1 - \cos\delta_\phi + \cos\delta_\theta) \end{cases} \quad (3.5)$$

Where $\delta_\theta$ is the phase of the empty wave and considering, as usual, that the two assumedly independent sources radiate with a random relative phase.

By setting the different arms of the interferometer with an equal length, we can impose:

$$\delta_\phi = \delta_\theta = 0 \quad (3.6)$$

And, by substitution in the previous expression, we get:

$$\begin{cases} I_1 = \frac{1}{2}I_0 \\ I_2 = \frac{1}{2}I_0 \end{cases} \quad (3.7)$$

Summarizing:

The predicted results for this experiment, assuming the real existence of empty pilot waves, the relative independence of the sources and no bosonic effect upon the empty waves, imply that there will be a change in the overall probability of detection, at the output ports of the interferometer. The expected difference, $\Delta = I_2 - I_1$, in the counting rate prediction of the two detectors C1 and C2 depends on the physical approach:

$$\text{Usual:} \quad \Delta^U = I_0$$

$$\text{de Broglie:} \quad \Delta^B = 0$$

An experimental improvement of the previous setup may be made using optical fibers, as indicated in the next drawing, Fig.3.2:

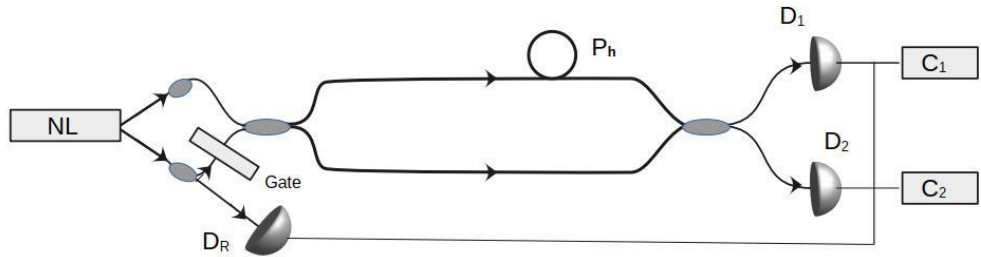

Fig.3.2 - Optical fiber Mach-Zehnder interferometer

As before, there is still the possible existence of a bi-photon effect that would render null the empty wave detection. We treat that difficulty next.

## 3.1 - Mach-Zehnder interferometric experiment variant to overcome the eventual bi-photon effect

As in the previous experimental setup, it is possible in the case of the Mach-Zehnder interferometer, to circumvent an eventual entanglement of the sources due to an eventual bi-photon effect [20]. We proceed as indicated in the Fig.3.3 below, resorting again to the use of two independent photon sources.

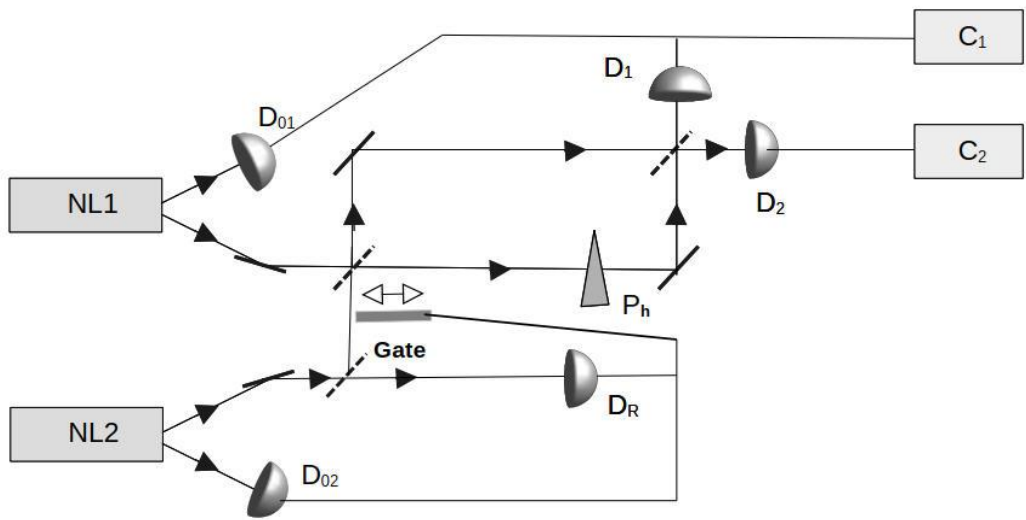

Fig.3.3 - Mach-Zehnder experiment variant to overcome the eventual bi-photon effect

The detectors $D_{01}$ and $D_{02}$ are placed in coincidence and used to ensure that NL1 and NL2 behave as monophotonic sources in the experience.

Since the two sources NL1 and NL2 are totally independent, the arrival of the photons does not happen in general at the same time but follows a random pattern. However, each time that a coincidence occurs, a change in the counting rate pattern is expected. In these conditions, a dip may then be seen in the steady continuous pattern, as shown schematically in Fig.3.4. Note that this figure is only schematic in the sense that what is important is the expected random appearance of dips, the size of which depends on the degree of overlapping of the waves. To increase the possibility of overlapping, it is convenient to raise the relative emission bit rate of the independent source NL2 relative to NL1.

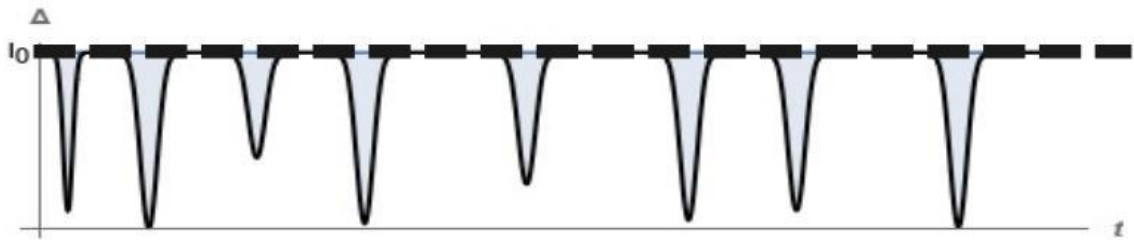

Fig.3.4 - Dotted line: usual predictions. Full line: sketch of expected behavior supposing the existence of de Broglie's empty waves

Again, this experimental setup may be improved with fiber optical technology, as shown in Fig.3.5:

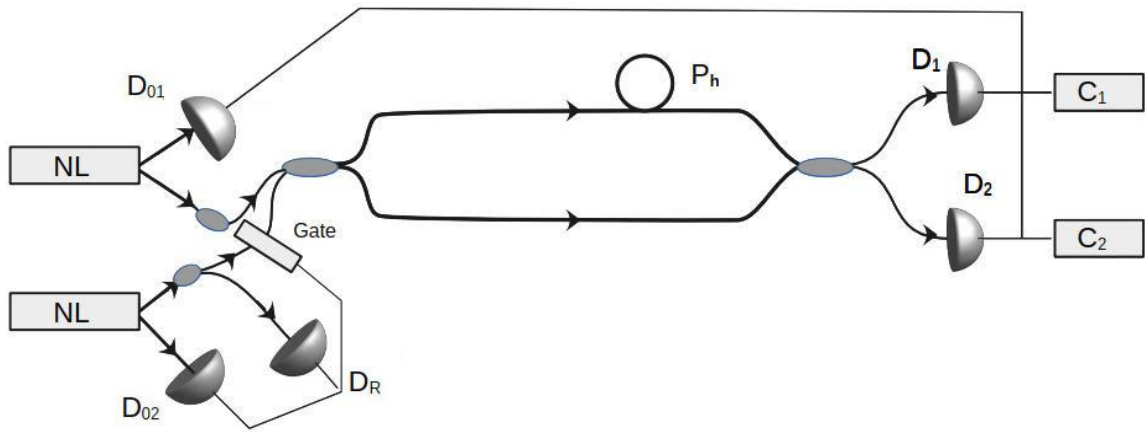

Fig.3.5 – Fiber optic Mach-Zehnder variant to overcome the eventual bi-photon effect

A variant of this experiment dealing specifically with the wave bosonic effect is presented next.

## 3.2 - Mach-Zehnder interferometric experiment variant to overcome the eventual wave bosonic effect

It is possible to design yet another experimental setup, now with a single photon source, that can also avoid the possibility of an eventual bosonic effect applying to quantum waves. A sketch of the device is shown in the next picture, Fig.3.6:

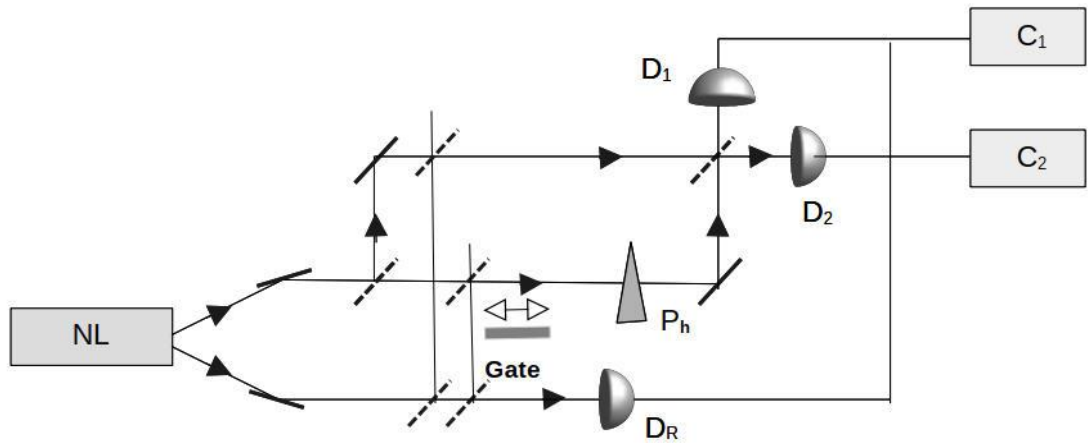

Fig.3.6 – Setup for overcoming the eventual bosonic effect applying to quantum waves

In this setup, to avoid the eventual bosonic effect, the empty guiding wave is not injected at the initial beamsplitter. The injection is made independently, along the upper and lower arm of the interferometer. In such conditions, the theta empty quantum waves mix at the second beamsplitter together with the other waves, with no particular phase correlation.

The calculations for this particular setup are practically the same as in the previous setup, shown in Fig.3.1. Consequently, when the gate is closed, blocking one of the upwards possible paths, the predictions of the two approaches, usual ("Copenhagen") method and realistic, de Broglie pilot wave hypothesis, are just the same. When the gate is open, the existence of the empty wave is manifested by the change in the steady counting rate, as sketched before in Fig.3.4.

An optical fiber variant of this setup is shown in next drawing, Fig.3.7:

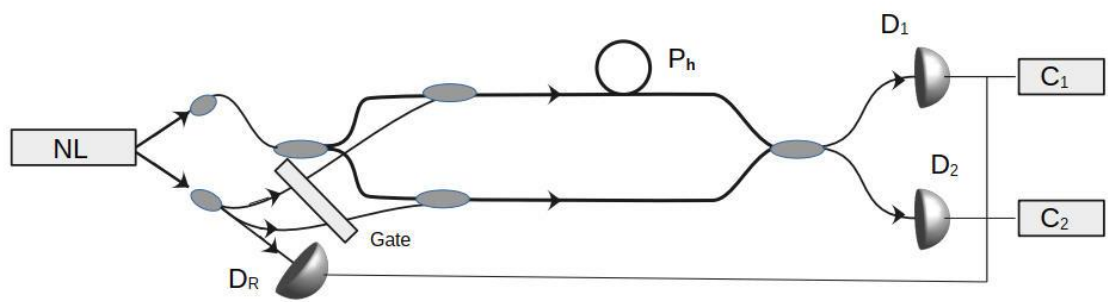

Fig.3.7 – Optical fiber variant to avoid the eventual wave bosonic effect

An experimental setup using Mach-Zehnder interferometry designed to deal with both the wave bosonic effect and the bi-photon effect is finally introduced next.

## 3.3 - Mach-Zehnder interferometric experiment variant to overcome both the eventual wave bosonic and biphoton effects

The process for overcoming the eventual bosonic and the entangled bi-photon effect follows along the same lines as before, using two independent sources, as shown in next picture, Fig.3.8:

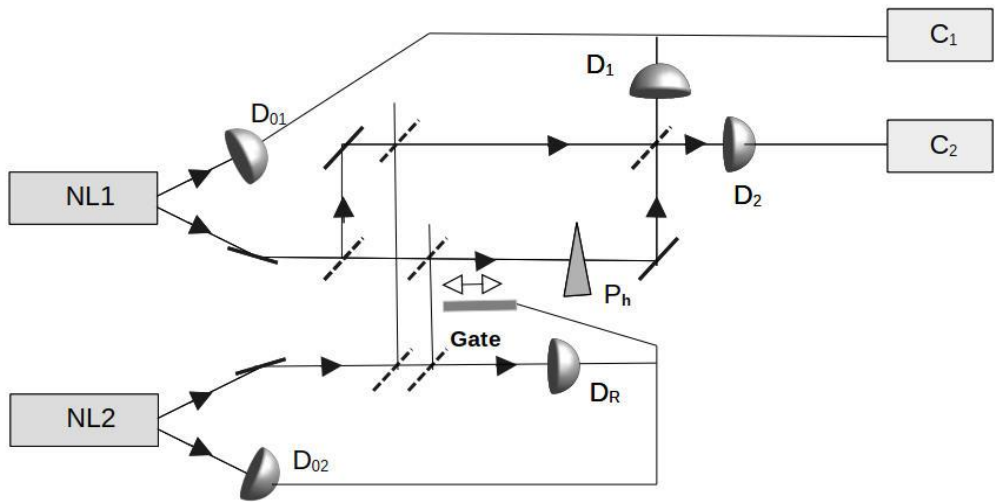

Fig.3.8 - Setup for overcoming the eventual bosonic and bi-photon effect

The expected results for this setup are essentially the ones of the device shown in Fig.3.3 above. Again, an optical fiber variant of this setup is possible, as shown in the next picture, Fig.3.9:

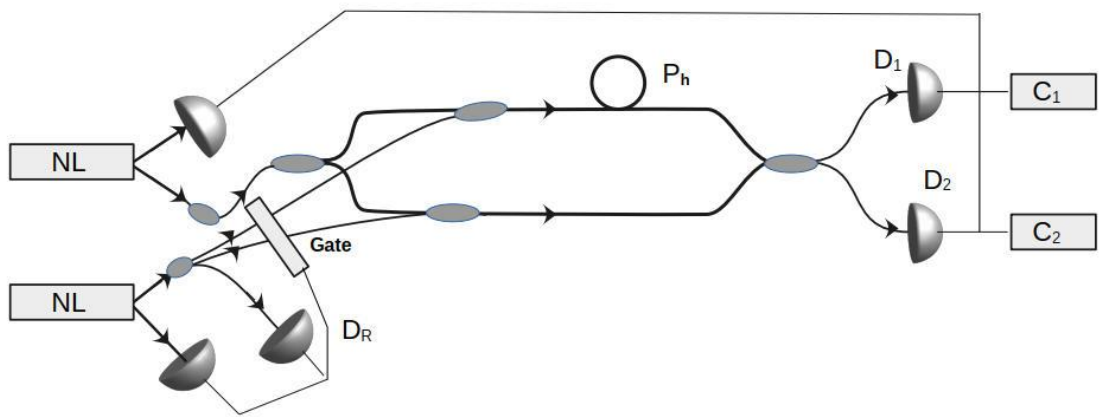

Fig.3.9 – Optical fiber setup for overcoming the eventual bosonic and bi-photon effects

## 4 – Neutron interferometry applied to the detection of empty quantum waves

Neutron interferometry has a long tradition in scientific research. Since neutrons are fermions, experiments designed to evaluate the existence of quantum waves, guiding neutrons, do not need, in principle, to take consider an eventual bosonic effect on the empty waves.

The major problem with these experiments [21] so far, has been the unavailability of sources capable of emitting pairs of neutrons at the same time. This fact has impeded the concrete realization of quantum empty wave detection experiments using neutrons.

However, this problem can be easily surpassed with an experimental design, analogous to the previous photonic setups, using independent uncorrelated sources. In such conditions, the experimental setup for the detection of neutron theta waves is shown in Fig.3.10 below.

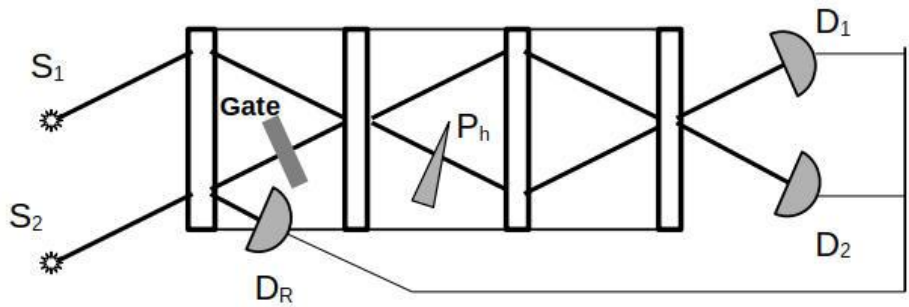

Fig.3.10 – Four slab neutron interferometer

As in Fig.2.2 above, here we also have two independent (now neutron) sources so that there is no chance of having correlations between the two quantum neutronic waves through some entanglement effect, similarly to what we did above for photons.

The gate is always closed except when the reduction collapse detector $D_R$ is triggered. In this situation we have two sources of neutrons that behave as if they were completely independent, with no phase correlations between the two neutrons. Consequently, the calculations for the expected experimental results follow much similarly to the ones made for the Mach-Zehnder interferometric experiment variant to overcome the eventual bi-photon effect, shown in Fig3.3 above.

The predictions for this neutron experiment when there occurs total wave superposition are:

a1) Usual ("Copenhagen") approach

$$\begin{cases} I_1^U = \frac{1}{2}I_0(1 - \cos \varphi_\phi) \\ I_2^U = \frac{1}{2}I_0(1 + \cos \varphi_\phi) \end{cases} \quad (4.1)$$

in which $\varphi_\phi$ is the constant phase difference between the traditional quantum waves.

a2) de Broglie's approach

$$\begin{cases} I_1^B = \frac{1}{2}I_0(1 - \cos\varphi_\phi + \cos\varphi_\theta) \\ I_2^B = \frac{1}{2}I_0(1 + \cos\varphi_\phi - \cos\varphi_\theta) \end{cases} \quad (4.2)$$

with $\varphi_\theta$ standing for the same constant phase shift of the two mixing theta waves.

By adequately setting the phase shift device $P_h$ (i.e., imposing $\varphi_\phi = \varphi_\theta = 0$), we have:

a2) Usual ("Copenhagen") approach:

$$\begin{cases} I_1^U = 0 \\ I_2^U = I_0 \end{cases} \quad (4.3)$$

that is, $\Delta^U = I_2^U - I_1^U = I_0$

b2) De Broglie's approach:

The predictions of the de Broglie approach depending on the degree of superposition.

b2.1) Total superposition:

$$\begin{cases} I_1^B = \frac{1}{2}I_0 \\ I_2^B = \frac{1}{2}I_0 \end{cases} \quad (4.4)$$

or

$$\Delta^B = I_2^B - I_1^B = 0. \quad (4.5)$$

b2.2) No superposition:

$$\Delta^B = I_2^B - I_1^B = \Delta^U = I_0. \quad (4.6)$$

Where the predictions are same as in the usual approach, $\Delta^B = \Delta^U$.

b2.3) Partial superposition:

In this case, there will be a change in the prediction for the counting difference at the two detectors, $\Delta^B = I_2^B - I_1^B$, varying from $I_0$ to 0 depending on the degree of overlapping, as indicated in Fig.3.4, above.

In summary, the evidence of the existence of neutron theta waves, or neutron empty waves, can be observed each time there is a change in the steady counting rate, as it was for the photonic counting rate.

**Conclusion**

We have proposed several experiments to detect quantum waves, feasible with present-day technologies, both from quantum optics and neutron interferometry. Our aim is to promote the experimental search for the interferometric evidence of real quantum empty waves. This, with the understanding, usual in de Broglie's realism tradition, that such waves are affected by splitting and reflection experimental conditions, just in the same way as quantum waves with corpuscles are. With this work we expect to have made some contribution to help clarify the ontic nature of quantum waves.

**Acknowledgements**

We would like to thank the anonymous referee for the comments and criticism which helped improve this presentation.

This work has been funded by FCT (Portugal) through CFCUL (project UID/FIL/00678/2019).

£